\documentclass[preprint,12pt]{elsarticle}
\pdfoutput=1

\usepackage[utf8]{inputenc}
\usepackage{graphicx}
\usepackage{amssymb}
\usepackage{amsmath}
\usepackage{color}
\usepackage{todonotes}
\usepackage{listings}
\usepackage{hyperref}
\usepackage{verbatim}
\usepackage{verbatimbox}
\usepackage{soul}
\usepackage{subcaption}
\usepackage{fancyvrb}
\usepackage[english]{babel}
\usepackage{braket}
\usepackage{hyperref}
\graphicspath{ {./images/} }

\newcounter{bla}

\definecolor{codegreen}{rgb}{0,0.6,0}
\definecolor{codegray}{rgb}{0.5,0.5,0.5}
\definecolor{codepurple}{rgb}{0.58,0,0.82}
\definecolor{backcolour}{rgb}{0.95,0.95,0.92}
 
\lstdefinestyle{mystyle}{
    backgroundcolor=\color{backcolour},   
    commentstyle=\color{codegreen},
    keywordstyle=\color{magenta},
    numberstyle=\tiny\color{codegray},
    stringstyle=\color{codepurple},
    basicstyle=\footnotesize,
    breakatwhitespace=false,         
    breaklines=true,                 
    captionpos=b,                    
    keepspaces=true,                 
    numbers=none,                    
    numbersep=5pt,                  
    showspaces=false,                
    showstringspaces=false,
    showtabs=false,                  
    tabsize=2
}
\journal{Computer Physics Communications}

\begin{document}
\begin{frontmatter}

\newcommand{\aldocomment} [1] 
{\todo[inline,backgroundcolor=green,size=\small ,bordercolor=white]{{\bf Aldo:} #1}}
\newcommand{\mattcomment} [1] 
{\todo[inline,backgroundcolor=red,size=\small ,bordercolor=white]{{\bf Francisco:} #1}}
\newcommand{\marcocomment} [1] 
{\todo[inline,backgroundcolor=orange,size=\small ,bordercolor=white]{{\bf Sobhit:} #1}}

\newcommand{\sscomment}[1]{\textcolor{magenta}{#1}}

\title{PyProcar: A Python library for electronic structure
  pre/post-processing}

\author[a]{Uthpala Herath}
\author[a]{Pedram Tavadze}
\author[b]{Xu He}
\author[b]{Eric Bousquet}
\author[a,c]{Sobhit Singh}
\author[d,e]{ Francisco Muñoz}
\author[a]{Aldo H. Romero\corref{author}}

\cortext[author] {Corresponding author.\\\textit{E-mail address:} alromero@mail.wvu.edu}
\address[a]{Department of Physics and Astronomy, West Virginia University, Morgantown, WV 26505-6315, USA}
\address[b]{Physique Th\'eorique des Mat\'eriaux, CESAM, Universit\'e de Li\`ege, B-4000 Sart-Tilman, Belgium}
\address[c]{Department of Physics and Astronomy, Rutgers University, Piscataway, NJ 08854, USA}
\address[d]{Departamento de F\'isica, Facultad de Ciencias, Universidad de Chile, Santiago, Chile}
\address[e]{Center for the Development of Nanoscience and Nanotechnology (CEDENNA), Santiago, Chile}

\begin{abstract}

The PyProcar Python package plots the band structure and the Fermi surface as a function of site and/or s,p,d,f - projected wavefunctions obtained for each $k$-point in the Brillouin zone and band in an electronic structure calculation. This can be performed on top of any electronic structure code, as long as the band and projection information is written in the PROCAR format, as done by the VASP and ABINIT codes. PyProcar can be easily modified to read other formats as well. This package is particularly suitable for understanding atomic effects into the band structure, Fermi surface, spin texture, etc. PyProcar can be conveniently used in a command line mode, where each one of the parameters define a plot property. In the case of Fermi-surfaces, the package is able to plot the surface with colors depending on other properties such as the electron velocity or spin projection. The mesh used to calculate the property does not need to be the same as the one used to obtain the Fermi surface. A file with a specific property evaluated for each $k$-point in a $k-$mesh and for each band can be used to project other properties such as electron-phonon mean path, Fermi velocity, electron effective mass, etc. Another existing feature refers to the band unfolding of supercell calculations into predefined unit cells. 

\end{abstract}


\end{frontmatter}
{\bf PROGRAM SUMMARY}

\begin{small}
\noindent
{\em Program Title:} PyProcar \\
{\em Licensing provisions:} GPLv3\\
{\em Programming language:} Python\\
{\em Supplementary material:}\\
{\em Nature of problem:} To automate, simplify and serialize the analysis of band structure and Fermi surface, especially for high throughput calculations. \\
{\em Solution method:} Implementation of a Python library able to handle, combine, parse,
extract, plot and even repair data from density functional
calculations. PyProcar uses color maps on the band structures or
Fermi surfaces to give a simple representation of the relevant
characteristics of the electronic structure.
{\em Additional comments:}\\ Features: PyProcar can produce high-quality figures
of band structures and Fermi surfaces (2D and 3D), projection of atomic orbitals, atoms, and/or spin components.\\
Restrictions: Only the VASP package is currently fully supported, the latest 
version of Abinit is partially supported (it will be fully supported in the the Abinit versions 9.x). The PROCAR file format can easily be implemented within  any DFT code. 
\end{small}

\section{Introduction}
\label{sec:intro}
\newcommand{\sscomment}[1]{\textcolor{magenta}{#1}}

Density Functional Theory (DFT) is one of the most widely used methodologies for electronic structure calculations in materials science \cite{harrison_introduction_nodate,PhysRev.140.A1133,PhysRev.136.B864}. Its very good quality/computational-cost ratio, together with the emergence of high performance computing, has reshaped the field of computational materials research. Decades efforts in developing DFT for simulation programs have led to highly efficient DFT codes which include, among others, Abinit \cite{gonze2016recent}, VASP \cite{PhysRevB.47.558, kresse_efficiency_1996}, Siesta \cite{Soler_2002} or Quantum Espresso \cite{QE-2017, QE-2009}, which are capable of exploring remarkable material properties. Thorough calculations with these DFT codes  have already been done for a wide range of compounds and the accumulated data can be publicly accessed from databases such as Materials Project \cite{Jain2013}, AFLOW \cite{curtarolo_aflow:_2012}, Materiae \cite{zhang_catalogue_2019}. This new approach to materials has had a large impact into condensed matter physics developing new paradigms such as the quantum materials~\cite{qm}. However, countless possibilities of exploring novel materials still exist and thus the necessity of efficient and reliable DFT pre- and post-processing tools.  

As computational capabilities increase, the size and complexity of the systems under study increase as well, hindering the analysis and the capability of abstraction on key aspects of these systems. As physicists, we encountered this problem and realized that we require a tool designed to gain insight from our calculations but spending less time on coding. In particular, we have three requirements for such a tool: (i) the generation of a graphical representation of the quantity of interest -surface states, spin-texture, orbital projection, etc.- should take only a single line of input. This approach also enables the scripting of similar analysis, a must for high throughput calculations. (ii) We also realized the need of producing high-quality graphics, ideally using a vector format, suitable for further post-processing. (iii) Finally, there are several issues preventing a straightforward analysis of electronic structure results, \textit{e.g.}, the size of output data files generated by DFT codes can be extremely large, thus making the post-processing of data very slow by excessive memory usage. In such a case, the post-processing tool, not the researcher, should be capable of taking care of large data files in a smart way and enabling the user to extract essential information while ignoring less interesting information from a large data file.

This paper focuses on the presentation of PyProcar, a robust,
open-source Python library used for pre- and post-processing of
the electronic structure data coming from DFT calculations. PyProcar is capable of performing a multitude of tasks including plotting plain and spin/atom/orbital projected band structures and Fermi surfaces- both in 2D and 3D, Fermi velocity plots, unfolding bands of a super cell, comparing band structures from multiple DFT calculations and generating a $k$-path for a given crystal structure. Our code has matured throughout the last decade and currently has over forty users who are members of the PyProcar forum \cite{PyProcarGoogleGroups-2019-05-29}. To retrieve a reliable number of users of an Open Source code is a difficult task due to inconsistencies in keeping track of downloads or installations, nevertheless, according to Google BigQuery \cite{GoogleBigQuery-2019-05-29} PyProcar has been downloaded almost 9000 times and PyPI Download Stats \cite{PyPIDownloadStats-2019-05-29} counts more than 100 installations of PyProcar per month based on the download analytics from the Python Packaging Index (PyPI) \cite{PyPIthePythonPackageIndexPyPI-2019-05-29}. Even though, PyProcar is written to parse the PROCAR file from the VASP package (and from
a fork of Abinit, which will be included in the next main production version), due to its object-oriented approach, PyProcar 
can be easily adapted to handle the output from other DFT codes. 
In Section \ref{sec:overview} we will briefly explain some basic 
aspects of PyProcar. Then in Section \ref{sec:features} we present some examples illustrating the capabilities of the PyProcar code for analysis of electronic structure. 


\section{Library Overview}
\label{sec:overview}

\subsection{Electronic structure projection}

The projection of the Kohn-Sham states over atomic orbitals and spins can give a large amount of information about the calculated system. However, such amount of information needs to be post-processed to extract the physical insight. In the VASP code, this information is written into the PROCAR file, which arranges the information of projections in blocks, as seen in Fig.~\ref{fig:procar} This block is repeated if calculations are spin polarized and for noncollinear spin calculations, three more blocks are present which correspond to $S_x$, $S_y$ and $S_z$ spin directions. There also exist a special format of PROCAR file that includes the phases of the wave function, as described in Fig.~\ref{fig:procar2}, which is an important quantity required for band unfolding. The direction of the atomic orbitals ($p_x$, $p_y$, $p_z$, etc.) is defined with respect to the Cartesian coordinates $x$, $y$ and $z$. 

\noindent
The site projected wave function in the PROCAR file is calculated by projecting the Kohn-Sham wave functions onto spherical harmonics that are non-zero within spheres of a Wigner-Seitz radius around each ion by: \break
\begin{equation*}
|\braket{Y^{\alpha}_{lm}|\phi_{nk}}|^2
\end{equation*}

\noindent 
where, $Y^{\alpha}_{lm}$ are the  spherical harmonics centered at ion index $\alpha$ with angular moment $l$ and magnetic quantum number $m$, and $\phi_{n \mathbf{k}}$ are the Kohn-Sham wave functions.  In general, for a non-collinear electronic structure calculation the same equation is generalized to:

\begin{equation*}
\frac{1}{2} \sum_{\alpha,\beta = 1}^2 \sigma_{\alpha,\beta}^i
\braket{\psi_{n,\mathbf{k}}^\alpha | Y_{lm}^{\alpha}} 
\braket{Y_{lm}^{\beta} | \psi_{n,\mathbf{k}}^\beta} 
\end{equation*}
where $\sigma^i$ are the Pauli matrices with $i = x, y , z$ and the spinor wavefunction $\phi_{n \mathbf{k}}$ is now defined as 
\begin{align*}
\phi_{n \mathbf{k}} & = \begin{bmatrix}
\psi_{n \mathbf{k}}^{\uparrow} \\
\psi_{n \mathbf{k}}^{\downarrow}
\end{bmatrix}
\end{align*}

\bigskip \noindent
This projection is performed for every $k$-point used in the DFT calculation, for every energy band and every atom of the crystal. Files can be stored as the absolute value of the projections, as it is usually needed for the band analysis or by also including the complex nature of the projections, which are used for the electron band folding. The raw data written on the PROCAR file captures most of the material details, essential to investigate interesting materials properties. But on the downside, these files are often too big to be analyzed without additional computational support. For instance, a spin non-polarized calculation involving 10 atoms, with 100 bands and 100 $k$-points gives almost $10^6$ entries in the PROCAR file, and the file size grows with the number of atoms as $\mathcal{O}(n^2)$. Therefore, it is essential to reduce the PROCAR file size in an efficient way without losing the key information stored in the file. The PyProcar parser can be used to manipulate, reduce, filter, and comprehend the information stored in the PROCAR file.

\begin{figure}
    \begin{verbnobox}[\fontsize{9pt}{9pt}\selectfont]
PROCAR lm decomposed
# of k-points:  160         # of bands:   50         # of ions:    5
    
 k-point     1 :    0.00000000 0.00000000 0.00000000     weight = 0.00625000
    
band     1 # energy  -29.05608270 # occ.  2.00000000
     
ion      s     py     pz     px    dxy    dyz    dz2    dxz  x2-y2    tot
    1  0.972  0.000  0.000  0.000  0.000  0.000  0.000  0.000  0.000  0.972
    2  0.000  0.000  0.000  0.000  0.000  0.000  0.000  0.000  0.000  0.000
    3  0.000  0.000  0.000  0.000  0.000  0.000  0.000  0.000  0.000  0.000
    4  0.000  0.000  0.000  0.000  0.000  0.000  0.000  0.000  0.000  0.000
    5  0.000  0.000  0.000  0.000  0.000  0.000  0.000  0.000  0.000  0.000
tot    0.974  0.000  0.000  0.000  0.000  0.000  0.000  0.000  0.000  0.974
     
band     2 # energy  -14.29383601 # occ.  2.00000000
     
ion      s     py     pz     px    dxy    dyz    dz2    dxz  x2-y2    tot
    1  0.011  0.000  0.000  0.000  0.000  0.000  0.000  0.000  0.000  0.011
    2  0.138  0.000  0.000  0.000  0.000  0.000  0.000  0.000  0.000  0.138
    3  0.234  0.000  0.000  0.000  0.000  0.000  0.000  0.000  0.000  0.234
    4  0.234  0.000  0.000  0.000  0.000  0.000  0.000  0.000  0.000  0.234
    5  0.234  0.000  0.000  0.000  0.000  0.000  0.000  0.000  0.000  0.234
tot    0.852  0.000  0.000  0.000  0.000  0.000  0.000  0.000  0.000  0.852
    \end{verbnobox}
    \caption{Example of the PROCAR file structure for a non spin polarized calculation. For spin polarized or noncollinear calculations there are additional blocks for each spin component. The PROCAR stores the real projections of the Kohn-Sham orbitals over the atomic orbitals in different blocks for each spin direction $(x,y,z)$, the total spin for each $k$-point in the Brillouin zone and for each electron band.}
  \label{fig:procar}
\end{figure}

\begin{figure}
    \begin{verbnobox}[\fontsize{9pt}{9pt}\selectfont]
PROCAR lm decomposed + phase
# of k-points:  160         # of bands:   50         # of ions:    5

 k-point     1 :    0.00000000 0.00000000 0.00000000     weight = 0.00625000

band     1 # energy  -29.05624010 # occ.  2.00000000
 
ion      s     py     pz     px    dxy    dyz    dz2    dxz  x2-y2    tot
    1  0.972  0.000  0.000  0.000  0.000  0.000  0.000  0.000  0.000  0.972
    2  0.000  0.000  0.000  0.000  0.000  0.000  0.000  0.000  0.000  0.000
    3  0.000  0.000  0.000  0.000  0.000  0.000  0.000  0.000  0.000  0.000
    4  0.000  0.000  0.000  0.000  0.000  0.000  0.000  0.000  0.000  0.000
    5  0.000  0.000  0.000  0.000  0.000  0.000  0.000  0.000  0.000  0.000
tot    0.974  0.000  0.000  0.000  0.000  0.000  0.000  0.000  0.000  0.974
ion          s             py             pz             px             dxy             
    1 -0.437 -0.883   0.000  0.000   0.000 -0.000  -0.000 -0.000   0.000 -0.000  ...
    2 -0.007 -0.015   0.000 -0.000   0.000 -0.000   0.000  0.000   0.000 -0.000  ...
    3 -0.009 -0.018  -0.000 -0.000   0.000  0.000  -0.000 -0.000   0.000  0.000  ... 
    4 -0.009 -0.018  -0.000  0.000   0.000 -0.000   0.000 -0.000   0.000  0.000  ... 
    5 -0.009 -0.018  -0.000 -0.000   0.000  0.000   0.000 -0.000   0.000  0.000  ... 
charge 0.973          0.000          0.000          0.000          0.000         ...
    \end{verbnobox}
    \caption{Example of the PROCAR file structure for a spin non-polarized calculation with a phase factor. This includes an additional block containing real and complex projections of the Kohn-Sham orbitals over the atomic orbitals.}
  \label{fig:procar2}
\end{figure}

\subsection{PyProcar Design}

PyProcar is an object-oriented code, and despite its name, only the file-related classes (mostly the parsing) are related to the VASP's PROCAR and OUTCAR. The OUTCAR gives a summary of the calculation including information about the electronic steps, eigenvalues, Fermi energy, forces on the atoms, \textit{etc}. PyProcar code parses the OUTCAR file and stores the Fermi energy, which is later used to shift the Fermi level to zero in band structure plots. The reciprocal lattice vectors are also parsed from OUTCAR. PyProcar is currently capable of parsing the outputs from  a recent version of Abinit and given the code's flexibility, the extension to other codes is trivial. 
Once the data is stored in memory, PyProcar can be used regardless of the employed DFT code. 
PyProcar consists of several classes and functions, as mentioned below. The functions provide a high-level interaction with the user, allowing the relevant task to get done with minimum instructions. For convenience, and to allow an easy scripting, all these functions are fully decoupled. Also, the manipulation of the data can be saved/exported as a new file on the disk.

The low level work is carried by classes, each handling one specific task or sub-task, they are fairly complex and no interaction with the user is expected unless a new feature is needed. By design, the classes are as loosely connected as possible. Each class uses the \texttt{logging} module to provide a user-defined level of verbosity, which is very useful for debugging.

\begin{figure}[ht!]
    \begin{center}
  \includegraphics[width=1.15\linewidth]{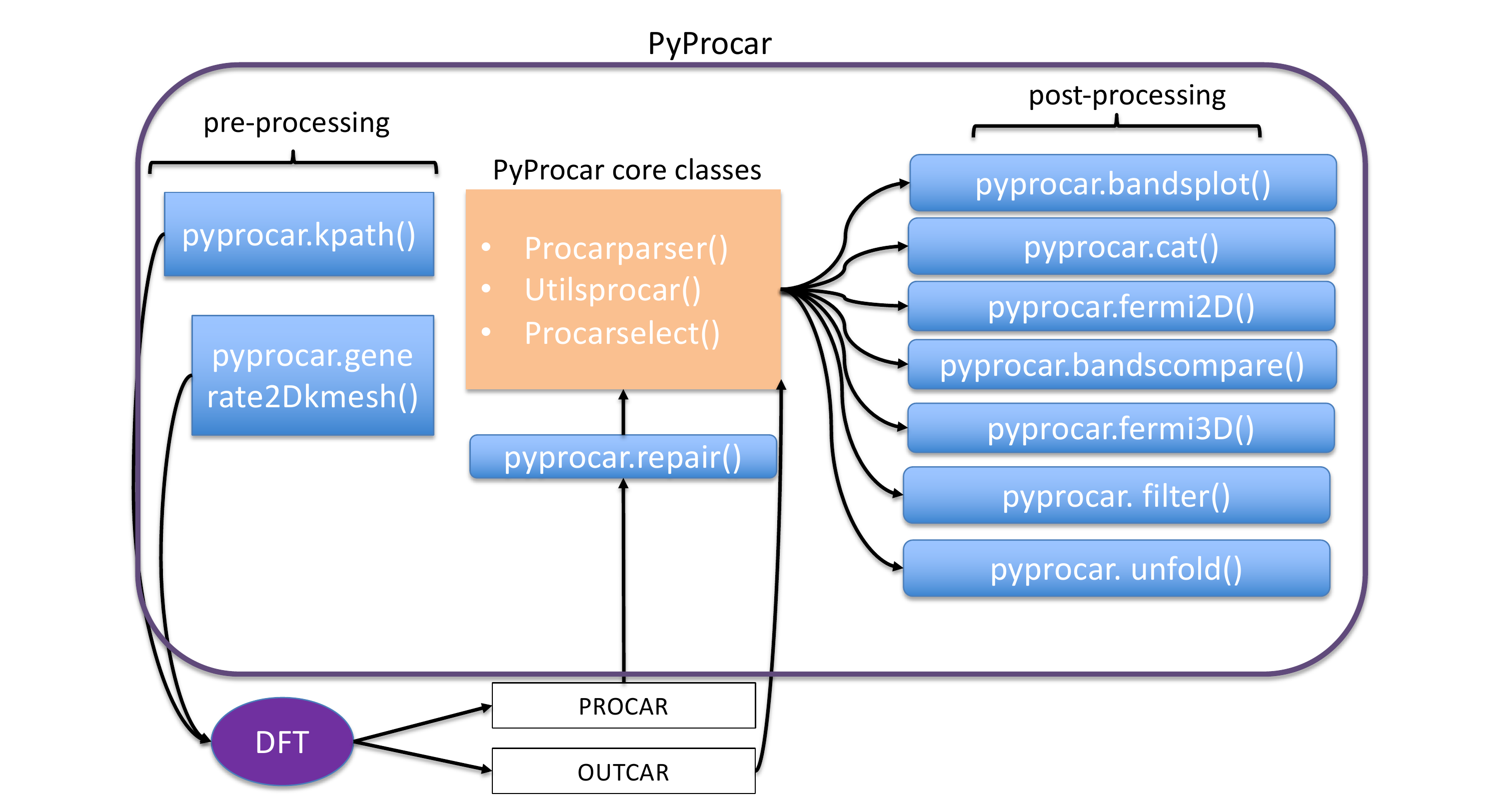}
  \caption{A structural overview of the PyProcar library}  
 \label{fig:scheme}
     \end{center}
\end{figure}

Fig.~\ref{fig:scheme} displays an overview of the PyProcar library. PyProcar can be used to generate files required for DFT calculations such as a suitable KPOINTS file for both self-consistent and non self-consistent DFT calculations. The structures can be generated manually, or from one of many databases publicly available such as Materials Project \cite{Jain2013}, AFLOW \cite{curtarolo_aflow:_2012}, etc. Once the DFT calculation is complete, PyProcar uses the VASP generated outputs, \texttt{PROCAR} and \texttt{OUTCAR}, for further post-processing.

The core classes within PyProcar consist of \texttt{Procarparser}, \texttt{Utilsprocar} and \texttt{Procarselect} which parses the \texttt{PROCAR} data and stores them in organized arrays for later calculations or analysis. These core classes are explained briefly below. 

\begin{enumerate}
  \item \texttt{Utilsprocar}:
	This class contains modules to parse the \texttt{OUTCAR} file from a DFT calculation. It reads and stores the Fermi energy and the reciprocal lattice vectors.  The Fermi energy is used to shift the Fermi level to zero in band structure calculations. The availability of the reciprocal lattice vectors enable \texttt{Utilsprocar} to convert the $k$-mesh grid between direct and Cartesian coordinates. The Fermi energy and reciprocal lattice vector could also be provided manually through command line inputs.

  \item \texttt{ProcarParser}:
    As the name suggests, this class contains modules required to parse the \texttt{PROCAR} file. It reads information related to bands, $k$-points and orbitals, and saves the data in the memory. It is designed to be somewhat resilient to errors derived from the fixed-format of the \texttt{PROCAR} file (\textit{i.e.,} missing blank spaces, use of \texttt{***} characters as an index, etc.)
  
  \item \texttt{ProcarSelect}
	After the \texttt{PROCAR} file is parsed and stored in memory, this class  manipulates the required orbital(s), atom(s) and or spin information separately.
\end{enumerate}

\subsection{Installation}

\noindent
The latest stable version of PyProcar, version 3.8.5 at the time of writing this paper, can be installed using the Python Packaging Index (\texttt{pip}) using the following command:
\begin{lstlisting}
pip install pyprocar
\end{lstlisting}
\noindent
The project's GitHub repository is located at \url{https://github.com/romerogroup/pyprocar}. An easy to follow documentation with examples can be found at \url{https://romerogroup.github.io/pyprocar/}.
PyProcar is supported by both Python 2.x and 3.x.

\section{Examples: MgB$_2$, BiSb, TaSb, and SrVO$_3$}
\label{sec:features}

\subsection{The MgB$_2$, BiSb, TaSb and SrVO$_3$ crystal structures}

\begin{figure}[ht!]
    \centering
    \includegraphics[width=\linewidth]{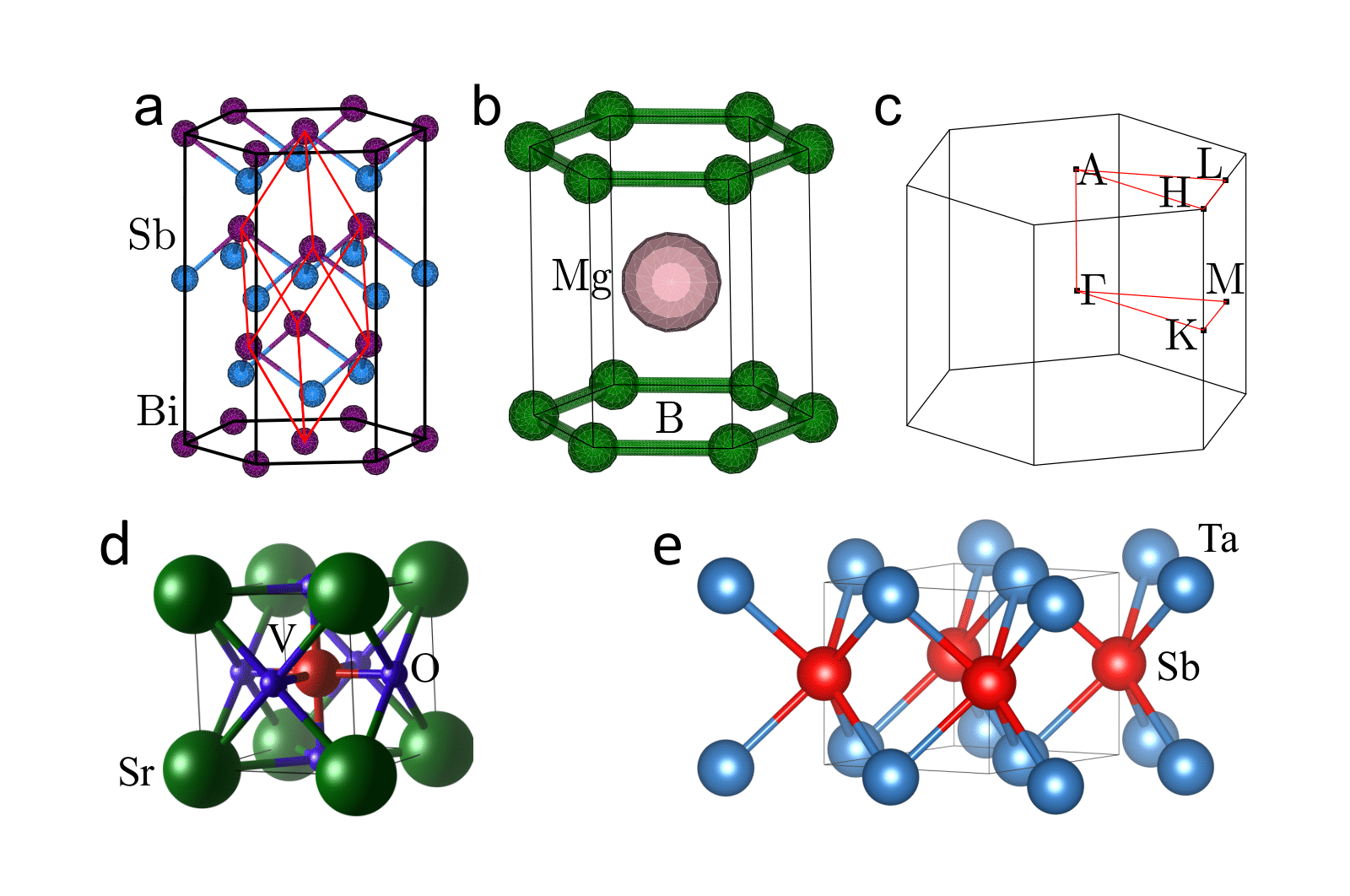}
    \caption{(a) Crystal structure of a phase of BiSb with rhombohedral symmetry, the primitive cell is enclosed by red lines. (b) Hexagonal lattice of MgB$_2$. (c) First Brillouin zone of a hexagonal lattice. (d) Cubic lattice of SrVO$_3$. (e) Hexagonal lattice of TaSb }
    \label{fig:crystal}
\end{figure}

In this section, we choose four example systems, namely MgB$_2$, BiSb, TaSb, and SrVO$_3$, to illustrate the capabilities of the PyProcar code. Fig.~\ref{fig:crystal} shows the crystal structure of these systems.
The DFT calculations were performed using the Vienna Ab Initio Simulation Package (VASP) version 5.4.4. Similar calculations can be performed using the forthcoming version of Abinit, version 9.x, where the keyword {\it prtprocar} needs to be added in the input. 

For the strongly correlated cubic perovskite SrVO$_3$ (crystal symmetry $Pm\bar{3}m$ and lattice constant 3.84 {\AA}), the plain and projected band structures were evaluated within the generalized gradient approximation (GGA) using PBE functional \cite{PhysRevLett.77.3865}. An 8 $\times$ 8 $\times$ 8 Monkhorst-Pack \cite{PhysRevB.13.5188} $k$-mesh and an energy cutoff of 600 eV were required for electron wavefunction convergence. We considered 10 valence electrons of Sr (3s$^2$3p$^6$4s$^2$), 5 valence electrons of V (3d$^3$4s$^2$)  and 6 valence electrons of O (2s$^2$2p$^4$) in the PAW pseudopotential.

The superconducting material MgB$_2$ of the crystal symmetry $P6/mmm$ and lattice parameters 3.07 {\AA} and 3.53 {\AA} for $a$ and $c$, respectively, was used to showcase the Fermi surface plotting and band unfolding capabilities of PyProcar. For the former, we used a 10 $\times$ 10 $\times$ 10 Monkhort-Pack grid with an energy cut-off of 700 eV within the GGA with the PBE exchange-correlation function. The latter was done with a 5 $\times$ 5 $\times$ 5 Monkhort-Pack grid with an energy cut-off of 500 eV within the GGA with the PBEsol exchange-correlation function~\cite{csonka2009assessing}. Both of these calculations were done with 2 valence electrons of Mg (3s$^2$) and 3 valence electrons of B (2s$^2$2p$^1$) in the PAW pseudopotential. 

To demonstrate the 2D spin texture plotting capability of PyProcar we use the Rashba semiconductor BiSb monolayer which belongs to the space group $P3m1$ with lattice parameters $a = b =$ 4.26\,{\AA}. The calculation, which included spin-orbit coupling (SOC) was performed with a $k$-mesh grid of 10 $\times$ 10 $\times$ 1 and an energy cut-off of 650 eV with PBE using 15 valence electrons of Bi (5d$^{10}$6s$^2$6p$^3$) and 5 valence electrons of Sb (5s$^2$5p$^3$) in the PAW pseudopotential.

PyProcar can also be used to investigate the band degeneracy and Dirac/Weyl points in topological materials. We use the material TaSb belonging to the space group $P\bar{6}m2$ with lattice parameters a = b = 3.58 {\AA}, c = 3.81 {\AA} to demonstrate this feature. An energy cut-off of 600 eV and $k$-mesh grid of 10 $\times$ 10 $\times$ 10 along with PBE and SOC were used for the calculation considering 5 valence electrons of Ta (5d$^3$6s$^2$) and 5 valence electrons of Sb (5s$^2$5p$^3$) in the PAW pseudopotential.

\subsection{Generating a $k$-path}

In order to plot a band structure, one must define a set of $k$-points following a desired $k$-path in momentum space. A common option is to employ the AFLOW software framework \cite{curtarolo_aflow:_2012}, the SeeK-path module in Materials Cloud \cite{Home-2019-05-31} or the  Bilbao Crystallographic Server \cite{Aroyo2011183,Aroyo:xo5013,articlebilbao} to generate the $k$-path. However, PyProcar's $k$-path generation utility enables a
the user to automatically generate a suitable and sufficient $k$-path given the crystal structure, typically read from the \texttt{POSCAR} file. 
The $k$-path generation utility within PyProcar is based on the Python library \texttt{seekpath} developed by Hinuma \textit{et al.} \cite{hinuma_band_2017}.  The red line in Fig.~\ref{fig:crystal} (c) depicts such a $k$-path of a hexagonal lattice generated using the algorithms of  \texttt{seekpath}. 

\bigskip
\noindent 
General format:  
\begin{lstlisting}
pyprocar.kpath(infile,grid_size,with_time_reversal,recipe,threshhold,symprec,angle_tolerence)
\end{lstlisting}

\noindent Usage:
\begin{lstlisting}
pyprocar.kpath(`POSCAR',40,True,`hpkot',1e-07,1e-05,-1.0)
\end{lstlisting}

\noindent More details regarding these parameters can be found in Ref. \cite{Pythonmodulesdocumentationseekpath184documentation-2018-10-04}.

\subsection{Repair}
This utility is used to repair the ill-formatting of the PROCAR file due to the erroneous file handling in Fortran,  particularly in a VASP calculation. This prevents issues arising from the lack of white space between a number and a negative sign, for instance \texttt{0.000000-0.5000000}. Typically, pyprocar.repair() is recommended to be applied before using any other utility. 

\bigskip
\noindent 
Usage:
\begin{lstlisting}
pyprocar.repair(`PROCAR',`PROCAR-repaired')
\end{lstlisting}

\subsection{$k$-mesh generator}
\label{subsec:kmeshgen}

This utility can be used to generate a 2D $k$-mesh centered at a given $k$-point and in a given $k$-plane. This is particularly useful in computing 2D spin-textures and plotting 2D Fermi surfaces. For example, the following command creates a 2D $k_{x}-k_{y}$ mesh centered at the $\Gamma$ point ($k_{z} = 0$) ranging from coordinates (-0.5, -0.5, 0.0) to (0.5, 0.5, 0.0) with a grid size of 0.02:

\bigskip
\noindent 
General format:  
\begin{lstlisting}
pyprocar.generate2dkmesh(x1,y1,x2,y2,grid_size)
\end{lstlisting}

Usage:
\begin{lstlisting}
pyprocar.generate2dkmesh(-0.5,-0.5,0.5,0.5,0.02)
\end{lstlisting}

\subsection{Band structure}
 PyProcar goes beyond the conventional plain band structure to plot the projected bands that carry even more information, which will be described shortly. The projected bands are color coded in an informative manner to portray fine details. PyProcar is capable of labeling the $k$-path names automatically, however, the user can manually input them as desired. This will be showcased in the next section. 


\begin{figure}[h!]
  \centering
  
  \begin{subfigure}[b]{0.45\linewidth}
    \includegraphics[width=\linewidth]{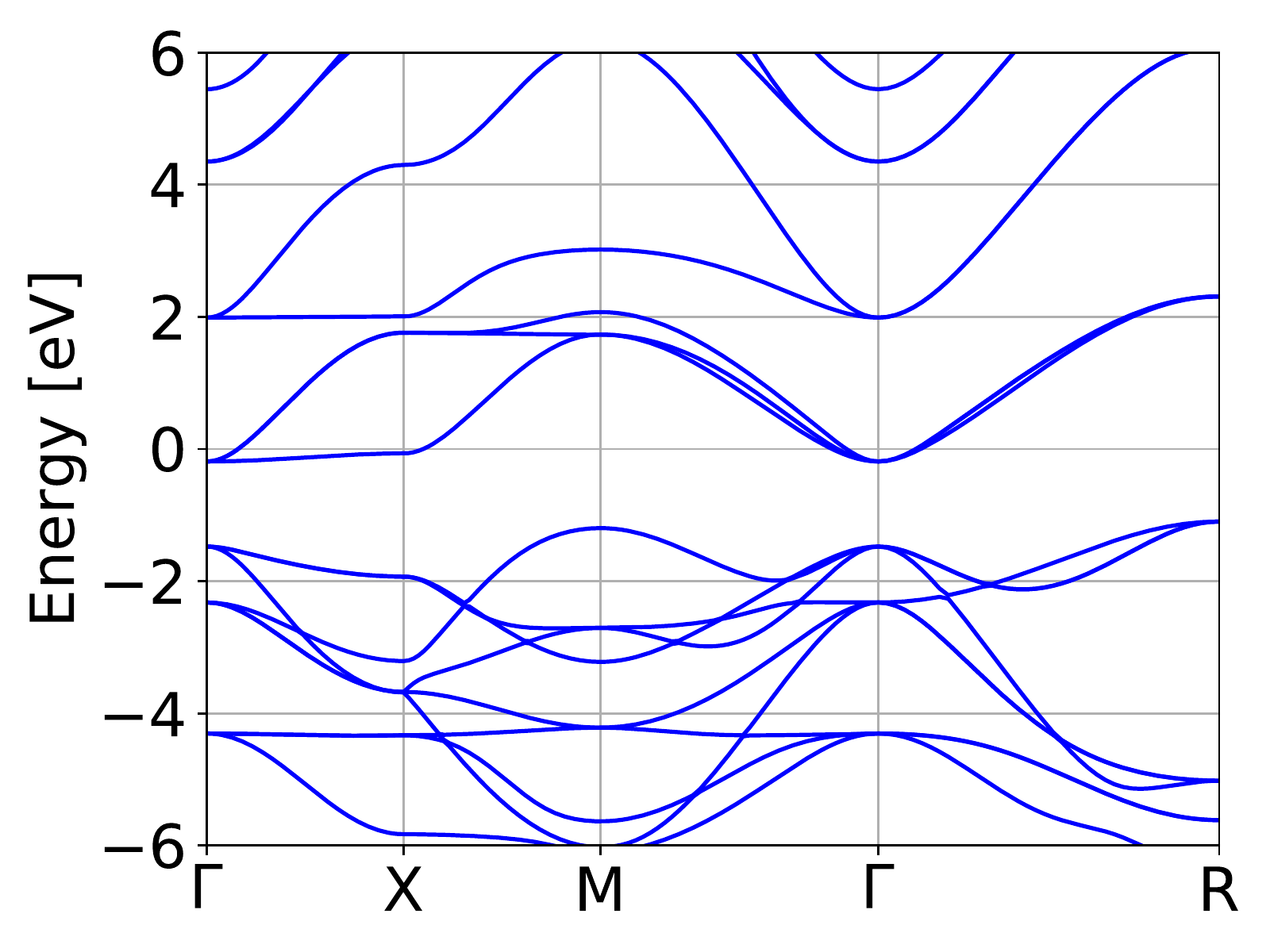}
     \caption{}
  \end{subfigure}
  \begin{subfigure}[b]{0.45\linewidth}
    \includegraphics[width=\linewidth]{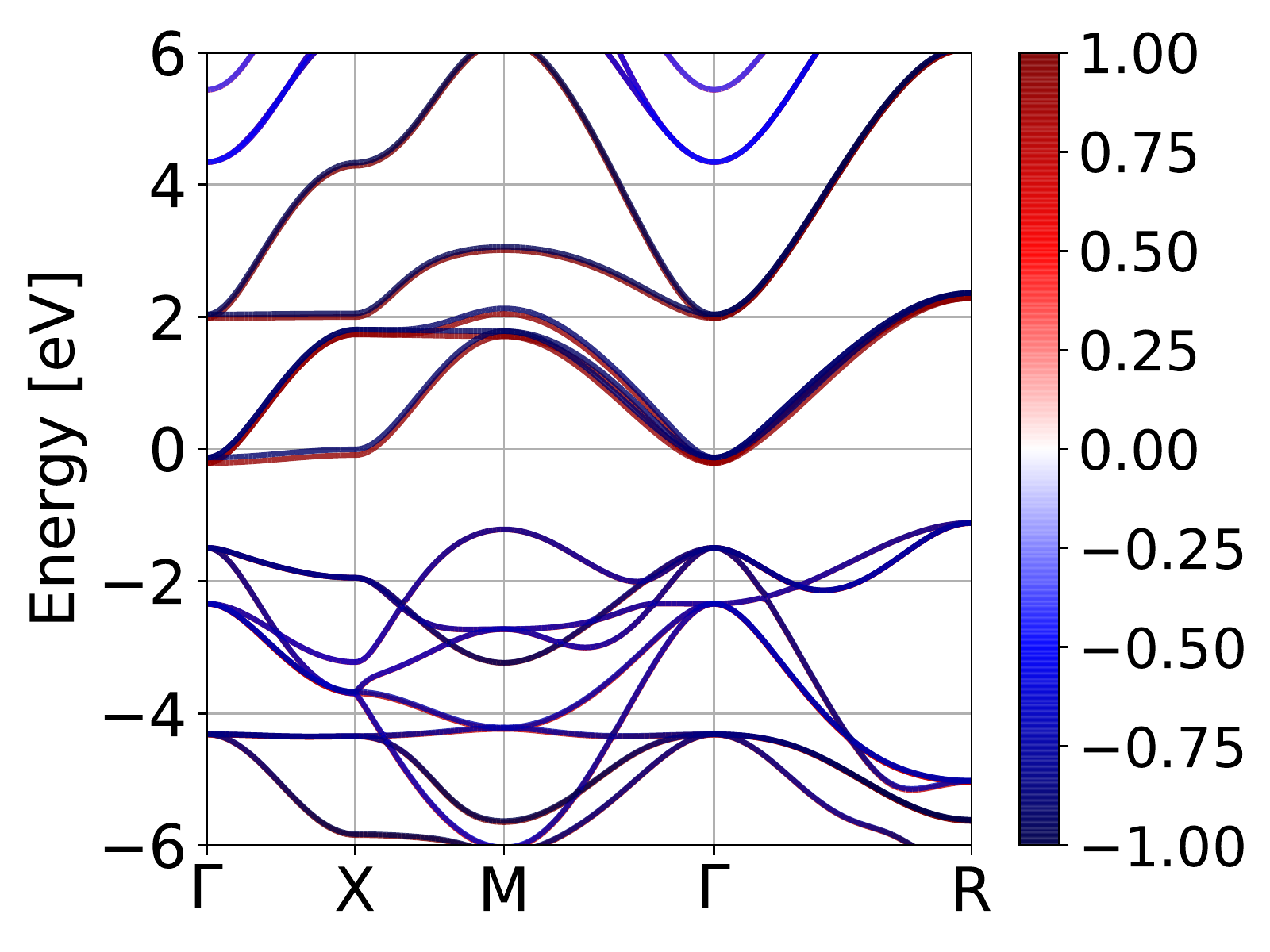}
    \caption{}
  \end{subfigure}
  \begin{subfigure}[b]{0.45\linewidth}
    \includegraphics[width=\linewidth]{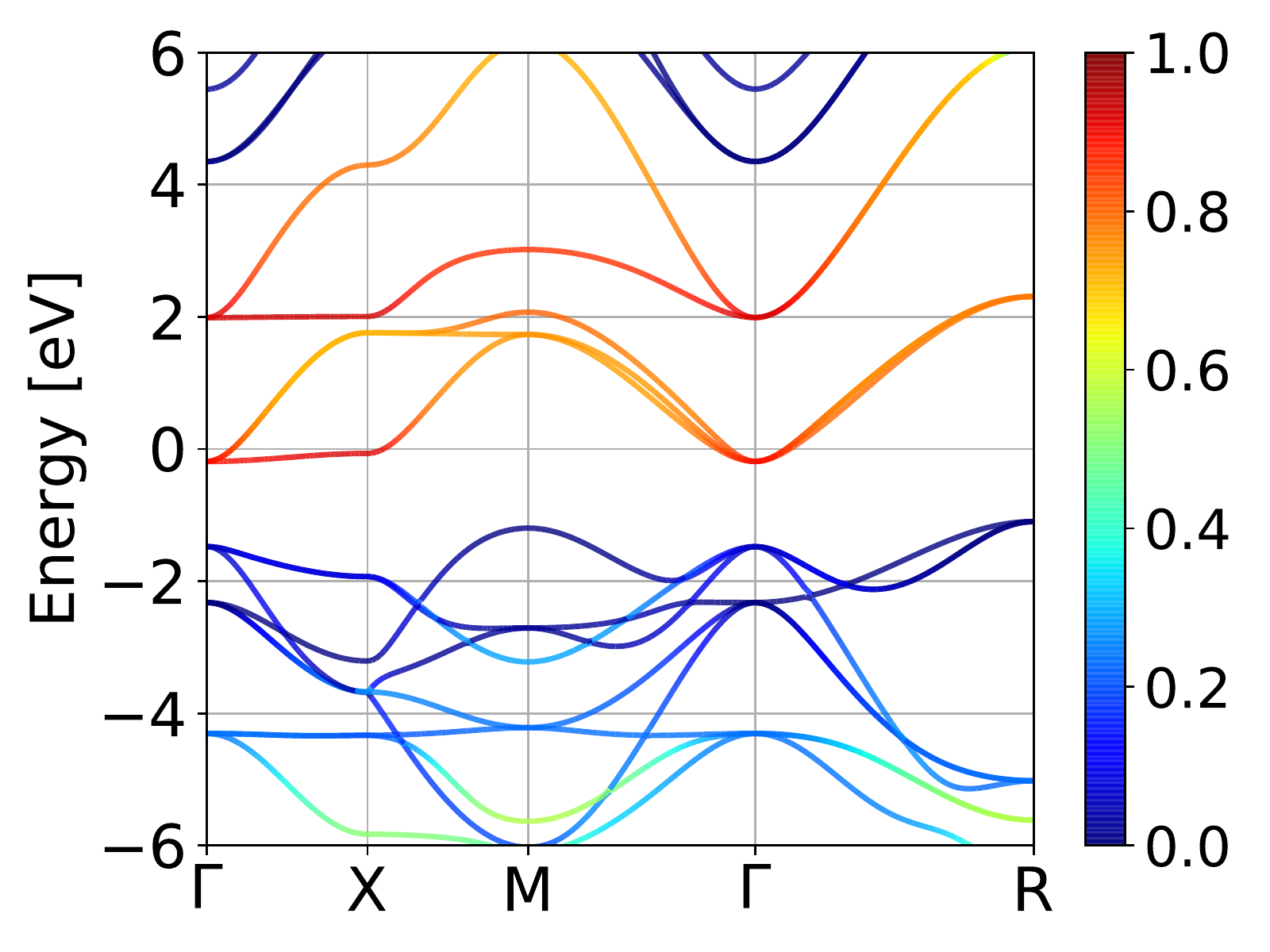}
    \caption{}
  \end{subfigure}
  \begin{subfigure}[b]{0.4523\linewidth}
    \includegraphics[width=\linewidth]{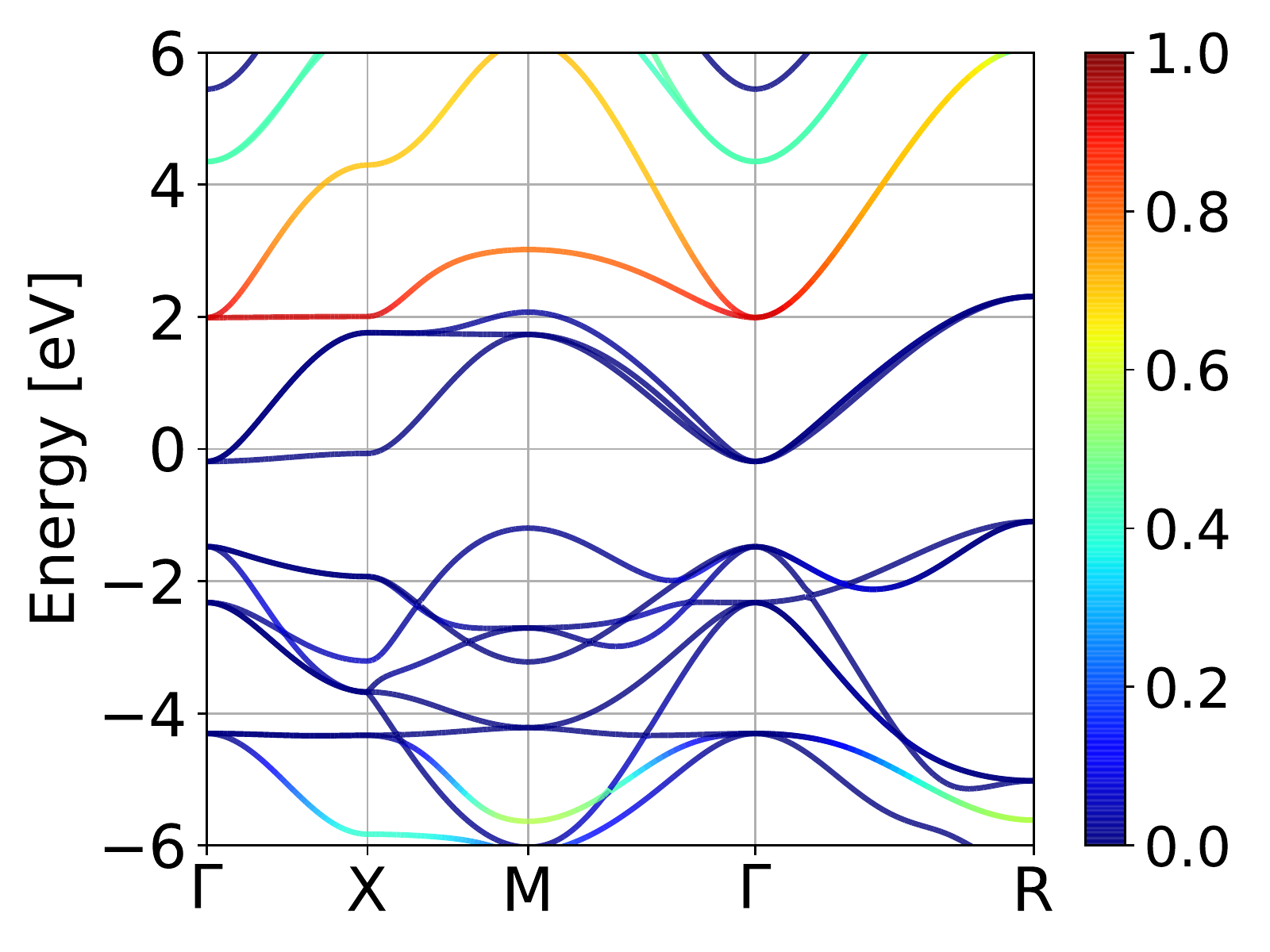}
    \caption{}
  \end{subfigure}
  
  \caption{(a) plain band structure (b) collinear spin projected (c) V atom projected (d) $e_{g}$ orbital projected band structure of SrVO$_3$. In the projected plots, the color intensity corresponds to the degree of contribution of that particular orbital, spin or atom type.}
  \label{fig:bandsplot}
\end{figure}

\begin{enumerate}

\item \textbf{No projection} \\
This is the most basic type of band structure. No projection information is contained here. See Fig.~\ref{fig:bandsplot} (a) for a plain band structure of SrVO$_3$. In order to use the plain mode one sets \textbf{mode}=`plain'. \textbf{elimit} sets the energy window limits. \textbf{outcar} specifies the \textbf{OUTCAR} file. For Abinit calculations, \textbf{abinit\_output} is used instead. \textbf{color} lets the user use any color available in the matplotlib \cite{Hunter:2007} package.   
\bigskip
\noindent 

Usage:
\begin{lstlisting}
pyprocar.bandsplot(`PROCAR-repaired',outcar=`OUTCAR',elimit=[-2,2],mode=`plain',color=`blue') 
\end{lstlisting}	
\bigskip

\item \textbf{Spin projection} \\
For collinear spin polarized and noncollinear spin calculations of DFT codes, PyProcar is able to plot the bands of each spin channel or direction separately. An example for a collinear spin polarized calculation is given in Fig.~\ref{fig:bandsplot} (b) where blue corresponds to spin down channel and red to spin up channel. For this case setting \textbf{spin}=`0' plots the total density from both spin channels and \textbf{spin}=`1' plots the spin channels separately. 

\bigskip
\noindent 

Usage:
\begin{lstlisting}
pyprocar.bandsplot(`PROCAR_repaired',outcar=`OUTCAR',elimit=[-5,5],kticks=[0,39,79,119,159],knames=[`G',`X',`M',`G',`R'],cmap=`seismic',
mode=`parametric',spin=`1')
\end{lstlisting}

For noncollinear spin calculations, spin=1,2,3 corresponds to spins oriented in S$_x$, S$_y$ and S$_z$ directions respectively. For parametric plots such as spin, atom and orbitals, the user should set \textbf{mode}=`parametric'. \textbf{knames} and \textbf{kticks} corresponds to the labels and the number of grid points between the high symmetry points in the $k$-path used for the band structure calculation. At the end of this section we explain how to retrieve this automatically. \LaTeX entries, such \verb|r`\Gamma'| also can be used. \textbf{cmap} refers to the matplotlib color map used for the parametric plotting and can be modified by using the same color maps used in matplotlib.

\bigskip

\item \textbf{Atom projection} \\
The projection of atoms onto bands can provide information such as which atoms contribute to the electronic states near the Fermi level. PyProcar counts each row of ions in the PROCAR file, starting from zero. In the example of a five atom SrVO$_3$, the indexes of \textbf{atoms} for Sr, V and the three O atoms would be 1,2 and 3,4,5 respectively.   It is also possible to include more than one type of atom. See Fig.~\ref{fig:bandsplot} (c) for an example of V atom projected bands in SrVO$_3$.
\bigskip
\noindent 

Usage:
\begin{lstlisting}
pyprocar.bandsplot(`PROCAR_repaired',outcar=`OUTCAR',elimit=[-5,5],kticks=[0,39,79,119,159],knames=[`G',`X',`M',`G',`R'],cmap=`seismic', mode=`parametric',atoms=[1])
\end{lstlisting}	  
\bigskip
  
\item \textbf{Orbital projection} \\
The projection of atomic orbitals onto bands is also useful to identify the contribution of orbitals to bands. For instance, to identify correlated $d$ or $f$ orbitals in a strongly correlated material near the Fermi level. It is possible to include more than one type of orbital projection. Fig.~\ref{fig:bandsplot} (d) displays an orbital projected band structure of SrVO$_3$. The mapping of the index of orbitals to be used in \textbf{orbitals} is as follows (this is the same order from the PROCAR file, see Fig.~\ref{fig:procar2}). 

\begin{center}
\scalebox{0.75}{
\begin{tabular}{ |c| c| c| c| c| c| c| c| c| c| c| c| c| c| c| c| }
\hline
 s & p$_y$ & p$_z$ & p$_x$ & d$_{xy}$ & d$_{yz}$ & d$_{z^2}$ & d$_{xz}$ & d$_{x^2-y^2}$ & f$_{y({3x^2-y^2})}$ & f$_{xyz}$  & f$_{yz^2}$ & f$_{z^3}$ &f$_{xz^2}$ & f$_{z({x^2-y^2})}$ & f$_{x({x^2-3y^2})}$  \\
\hline 
0 & 1 & 2 & 3 & 4 & 5 & 6 & 7 & 8 & 9 & 10 &11 & 12 & 13 & 14 & 15   \\
\hline
\end{tabular}}
\end{center}

\bigskip
\noindent 

Usage:
To project all five $d$-orbitals: 
\begin{lstlisting}
pyprocar.bandsplot(`PROCAR_repaired',outcar=`OUTCAR',elimit=[-5,5],kticks=[0,39,79,119,159],knames=[`G',`X',`M',`G',`R'],cmap=`seismic',mode=`parametric',orbitals=[4,5,6,7,8])
\end{lstlisting}	    
\end{enumerate}

\noindent If a KPOINTS file from a VASP calculation is present, PyProcar automatically retrieves \textbf{kticks} and \textbf{knames} and labels the band structure plots accordingly for any of the above cases and also for the \texttt{bandscompare()} function in section \ref{sec:comparebands}. For example:
\begin{lstlisting}
pyprocar.bandsplot(`PROCAR_repaired',outcar=`OUTCAR',mode='plain',kpointsfile=`KPOINTS')
\end{lstlisting}

\noindent One or many of the above can be combined together to allow the user to probe into more specific queries such as a collinear spin projection of a certain orbital of a certain atom. 

Different modes of band structures are useful for obtaining information for different cases. The four modes available within PyProcar are \texttt{plain, scatter, parametric} and \texttt{atomic}. The \texttt{plain} bands contain no projection information. The \texttt{scatter} mode creates a scatter plot of points. The \texttt{parametric} mode interpolates between points to create bands which are also projectable. Finally, the \texttt{atomic} mode is useful to plot energy levels for atoms. 

These projected band structures play a vital role in revealing the physics of a system. For instance, the V$^{4+}$ projected bands displayed in Fig.~\ref{fig:bandsplot} (c) correctly captures the contribution from the V-$3d$ electrons with O-$2p$ states from -6 eV to -1 eV . Furthermore, in Fig.~\ref{fig:bandsplot} (d) where $e_g$ states are projected, one can clearly notice the octahedral crystal field splitting in SrVO$_3$ where $t_{2g}$ states exist below $e_g$ states, as expected.

\subsection{3D Fermi surface}
PyProcar's 3D Fermi surface utility is able to generate Fermi surface plots projected over spin, atoms and orbitals or a combination of one or many of each. This utility is also capable of projecting external properties that are provided on a mesh grid in momentum space. This feature is useful when one wants to project properties that are not provided in a PROCAR file such as Fermi velocity, electron-phonon coupling and electron effective mass. We divide this section into three sub sections, plain Fermi surface, projection of properties from PROCAR and projection of properties from external file.  

\begin{enumerate} 
\item  \textbf{Plain Fermi surface} \\
The \texttt{ProcarParser} class provides eigenvalues on a momentum space mesh grid. In all of the 3D Fermi surface functions in PyProcar the eigenvalues are interpolated using Fourier transform. This type of interpolation is suitable for eigenvalues as they are periodic in nature. The points in the mesh grid is generated in reduced space and some of the points sampled by VASP might not be in the first Brillouin Zone (BZ). The points outside of the first BZ can be returned to the first BZ using the reciprocal lattice vectors. As transforming the points to the first BZ will cause the distortion in the shape of the iso-surface, it is better to add the points missing from the first BZ then remove the points lying in the second BZ. In order to identify these points, we first create the boundaries of the first BZ by creating the Wigner-Seitz cell from the reciprocal lattice vectors. The BZ is a convex hull created by the points located at the corner of the BZ. To figure out if a point is inside the first BZ or outside the following procedure is performed on each point. A point is added to the collection of the points on the BZ corners and a new convex hull is created, if the convex hull is the same as the original BZ, the point is located inside, and if the BZ is different, the point is located outside of the first BZ. Since this operation can be slow in a scripting language like Python, we use data parallelism. This task is offloaded to a number of worker processes, which can be defined by the user. After obtaining enough points we use Lewiner marching cubes algorithm provided by scikit-image package\cite{scikit-image}. This function provides vertices and faces required to create the Fermi surface. The last step that is needed is to transform the coordinates of the vertices from the reduced coordinates to the Cartesian coordinates. To visualize the Fermi surface, we have provided four different plotting packages, mayavi\cite{mayavi}, matplotlib\cite{matplotlib}, plotly\cite{plotly} and ipyvolume \cite{maartenbreddelsipyvolume3dplottingforPythonintheJupyternotebookbasedonIPythonwidgetsusingWebGL-2019-06-06} which can be chosen by the user. The figures shown here are produced using mayavi. 

Usage:
\begin{lstlisting}
pyprocar.fermi3D(procar,outcar,bands,scale=1,mode=`plain',st=False,**kwargs)
\end{lstlisting}
The main arguments in this function are \textbf{procar}, \textbf{outcar}, \textbf{bands}, \textbf{scale}, \textbf{mode} and \textbf{st}, where \textbf{procar} and \textbf{outcar} are the names of the input PROCAR and OUTCAR files respectively, \textbf{bands} is an array of the bands that are desired to be plotted. Note if \textbf{bands} = -1, the function will try to plot all the bands provided in the PROCAR file. The $k$-mesh will be interpolated by a factor of \textbf{scale} in each direction. The \textbf{st} tag controls the spin-texture plotting, and \textbf{mode} determines the type of projection of colors. There are additional keyword arguments that can be accessed in the help section of this function, such as \textbf{face\_color}, \textbf{cmap}, \textbf{atoms}, \textbf{orbitals}, \textbf{energy}, \textbf{transparent}, \textbf{nprocess} \textit{etc.} Fig.~\ref{fig:plain_fermi3D} shows the Fermi surface of MgB$_2$ generated using \texttt{`plain'} mode of this function.


\begin{figure}[h!]
  \centering
  
  \begin{subfigure}[b]{0.4\linewidth}
    \includegraphics[width=\linewidth]{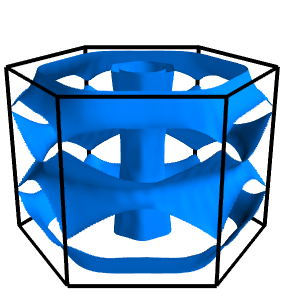}
     \caption{}
  \end{subfigure}\qquad
  \begin{subfigure}[b]{0.41\linewidth}
    \includegraphics[width=\linewidth]{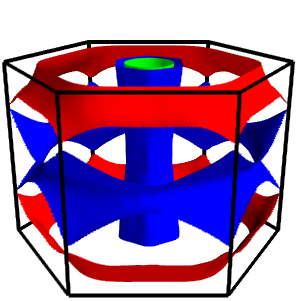}
    \caption{}
  \end{subfigure}

  \caption{(a) Plain 3D Fermi surface of MgB$_2$, (b) Plain 3D Fermi surface of MgB$_2$ with \textbf{face\_colors} specified. Note that the face colors have to be provided in tuples of (r,g,b) normalized to 1.}
  \label{fig:plain_fermi3D}
\end{figure}

\item  \textbf{Surface coloring based on properties from PROCAR}

Similar to the \texttt{bandsplot} section one can choose to project the contribution of different properties provided in the PROCAR file, such as atom, orbital and spin contributions. The projection can be represented by different color mapping schemes chosen by the user. The projection is not restricted to only one property at a time, so it can be chosen from all the provided properties. For example, one might want to see the contribution of the orbitals p$_x$, p$_y$, p$_z$ from specific atoms, this function will parse the desired contributions and projects the sum of contributions on each face. To do so, we perform an interpolation on the data provided by the DFT package and evaluate the function at the center of each face created in the previous section. To use this functionality one has to change the \textbf{mode} from \texttt{`plain'} to \texttt{`parametric'} and choose the atoms, orbitals, spin that are desired to be projected. As an example, we project the different contribution of different p orbitals of Boron atom in Fig.~\ref{fig:parametric_fermi3D}. The results show the middle cylinder is mostly comprised of p$_x$ and p$_y$ orbitals while the bands closer to the edges of BZ are p$_z$ orbitals. This is in agreement with the calculations performed using the SIESTA DFT package~\cite{MgB2projection}. 

\begin{figure}[h!]
  \centering
  
  \begin{subfigure}[b]{0.4\linewidth}
    \includegraphics[width=\linewidth]{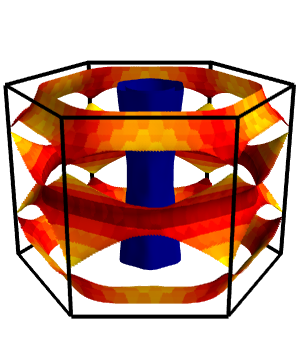}
     \caption{}
  \end{subfigure}\quad
  \begin{subfigure}[b]{0.4\linewidth}
    \includegraphics[width=\linewidth]{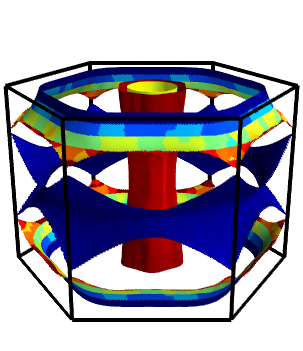}
    \caption{}
  \end{subfigure}\quad \hspace{0.1pt}
    \begin{subfigure}[b]{0.12\linewidth}
    \includegraphics[width=\linewidth]{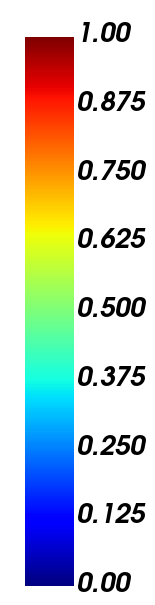}
  \end{subfigure}
  \caption{(a) The projection of p$_z$ orbitals of Boron atoms on Fermi surface of MgB$_2$ (b) Projection of p$_x$, p$_y$ orbitals of Boron atoms on Fermi surface of MgB$_2$
  }
  \label{fig:parametric_fermi3D}
\end{figure}

For noncolinear calculations, this function is able to plot arrows in the direction of the spinors provided in the PROCAR file. To turn this functionality on the one can set \texttt{\textbf{st}=True} to turn the spin-texture ON. The user can choose between coloring all the arrows originated from one band with a specific color, or project the contribution of that arrow in a specific Cartesian direction. We plot two examples to demonstrate this functionality in Fig.~\ref{fig:spin_texture_fermi3D}, the spin-texture of BiSb at 0.60\,eV above the Fermi level and the spin-texture of SrVO$_3$. To better represent the spin-texture we use the key argument \texttt{\textbf{transparent}=True} which changes the opacity of the Fermi surface to zero.

\begin{figure}[h!]
  \centering
  
  \begin{subfigure}[b]{0.4\linewidth}
    \includegraphics[width=\linewidth]{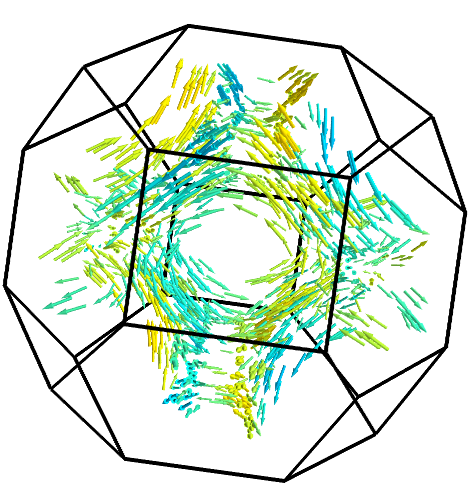}
     \caption{}
  \end{subfigure}\qquad
  \begin{subfigure}[b]{0.35\linewidth}
    \includegraphics[width=\linewidth]{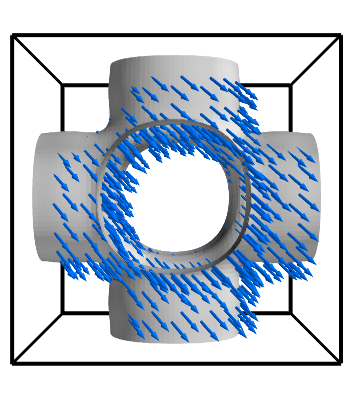}
    \caption{}
  \end{subfigure}\qquad
    \begin{subfigure}[b]{0.125\linewidth}
    \includegraphics[width=\linewidth]{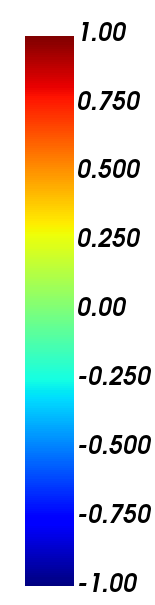}
  \end{subfigure}
  \caption{(a) The spin-texture of BiSb calculated above the Fermi-level at $E = E_{F} + 0.60$\,eV. One can clearly notice Rashba-type spin-splitting of conduction band electrons by analysing spin-texture at \textbf{k} and \textbf{-k} wave vectors. (b) Spin texture of  SrVO$_3$.}
  \label{fig:spin_texture_fermi3D}
\end{figure}

\item  \textbf{Surface coloring based on properties obtained from an external file}

Similar to the previous section, this function is able to read an external file, containing information about a scalar or a vector field in BZ and project the field on the Fermi surface. This file does not need to have the same mesh grid as the PROCAR file as long as the mesh sampling is fine enough. This function performs an interpolation on the provided data and evaluates functions at the center of each face on the Fermi surface. The external file should have the following format. 

   \begin{verbnobox}[\fontsize{10pt}{10pt}\selectfont]
band = <band number>
   <kx1>  <ky1>  <kz1>  <color1>
   <kx2>  <ky2>  <kz2>  <color2>
   <kx3>  <ky3>  <kz3>  <color3>
   ...
band = <band number>
...
    \end{verbnobox}
    
The function matches information about the first band present in the file to the first band requested to be plotted, second band present in the file to the second band requested to be plotted, and so on. As an example, we have plotted the Fermi velocity of MgB$_2$ calculated by numerically evaluating the energy gradient from the PROCAR file in Fig.~\ref{fig:external_fermi3D}. Of course, one can acquire more accurate Fermi velocity by means of the Wannier interpolation method \cite{marzari2012maximally}, however the following plot shows close enough accuracy for an example designed to represent the capabilities of this functional mode.

\begin{figure}[h!]
  \centering

  \begin{subfigure}[b]{0.6\linewidth}
    \includegraphics[width=\linewidth]{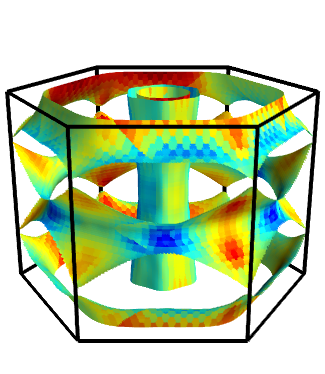}
  \end{subfigure}\qquad\quad
  \begin{subfigure}[b]{0.19\linewidth}
    \includegraphics[width=\linewidth]{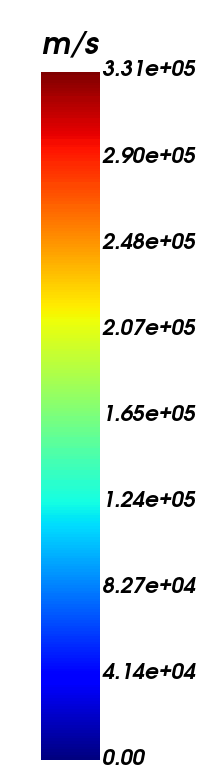}

  \end{subfigure}
  
  \caption{Fermi velocity of MgB$_2$ projected on the Fermi surface.}
  \label{fig:fermi_velocity_fermi3D}
  
  \label{fig:external_fermi3D}
\end{figure}

\end{enumerate}

\subsection{Handling big files: filtering the selected data and reducing the memory requirement for post-processing}
A simpler version of a PROCAR file containing only a subset of information from the original dataset can be generated with this utility.  This feature is very useful when there are many bands in the PROCAR file (e.g. in heterostructures or supercells calculations) making the file size enormously large for post-processing while only bands near the Fermi-level are needed for analysis. In this case, one can filter data of selected bands near the Fermi-level. This considerably reduces the file size and makes the post-processing of data faster. For instance, in graphene/MoS$_2$ (5:4) heterostructures~\cite{grMoS2_PRB2018, singh2018proximity} the size of a VASP generated PROCAR file is 1.49 GB (98 atoms, 584 electrons, 804 bands in total, 63 $k$-points, and spin-orbit coupling included). However, by filtering only 30 bands above and below the Fermi-level ($i.e.$ band indexes 554--614), the file size reduces to 108 MB and data processing becomes much faster. In the same way, one could use the \texttt{filter} utility to filter the \texttt{PROCAR} file to extract information regarding particular spins, atoms, or orbitals in a relatively smaller \texttt{PROCAR-new} file.

The following example extracts information of bands ranging from index 50 to 70 from a \texttt{PROCAR-repaired} file (Fermi-level is near band \#60) while ignoring all other bands located far from the Fermi-level, and stores resulting dataset in a new file named \texttt{PROCAR-repaired-band50-70}. Now the new \texttt{PROCAR-repaired-band50-70} file can be used for further post-processing of data at relatively low memory requirements.

\bigskip
\noindent 
Usage:
\begin{lstlisting}
pyprocar.filter(`PROCAR-repaired',`PROCAR-repaired-band50-70',bands=[50,70])
\end{lstlisting}


\subsection{2D spin-texture}
This module can be utilized to visualize the constant energy surface spin textures in a given system. This feature is particularly useful in identifying Rashba and Dresselhaus type spin-splitting effects, analyzing the topology of Fermi surface, and examining Lifshitz transitions, as demonstrated in Refs.~\cite{BiSbPCCP2016, singhWeyl2016, CamiloPRB2016, singhPRB2017_Rashba, singh2018proximity, StructuralPredictionSOC_2018, monolayerBi_2019}. To plot 2D spin texture, we require a 2D $k$-grid centered a certain special $k$-point in Brillouin zone near which we want to examine the spin-texture in $k$-space (see section~\ref{subsec:kmeshgen} regarding generation of 2D $k$-mesh). 

\bigskip
\noindent 
Usage:
\small{To plot $S_x$ spin component at a constant energy surface $E = E_{F} + 0.60\,eV$ (spin=`1', `2', `3' for $S_x$, $S_y$, $S_z$, respectively)}
\begin{lstlisting}
pyprocar.fermi2D(`PROCAR-repaired', outcar=`OUTCAR', st=True, energy=0.60, noarrow=True, spin=`1')
\end{lstlisting}	 

\begin{figure}[htpb!]
  \includegraphics[width=14cm, keepaspectratio=true]{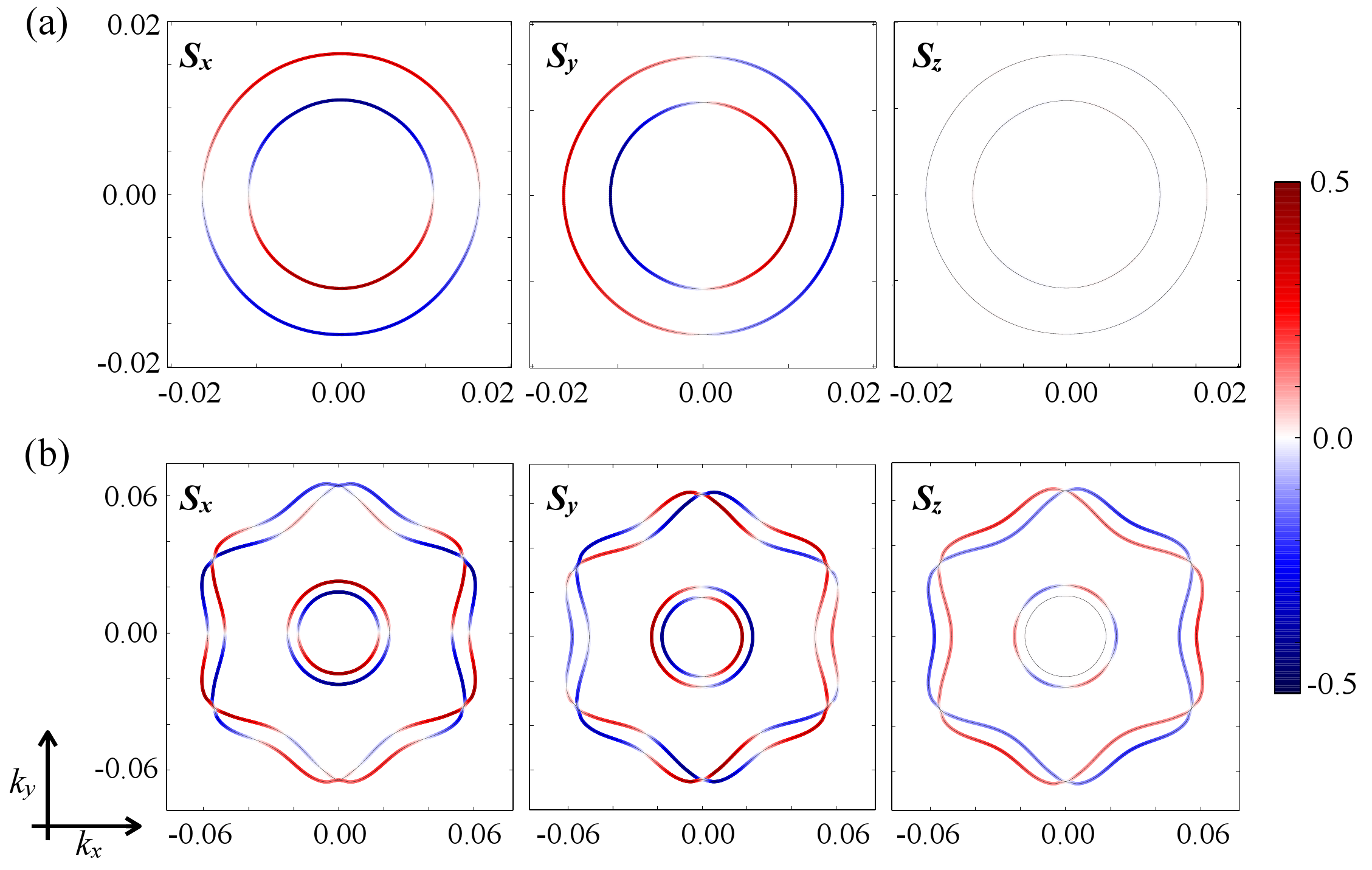}
  \caption{Spin texture in BiSb monolayer, calculated at a constant energy surface, above and below the Fermi-level, respectively, at (a) $E = E_{F} + 0.60\,eV$, and (b) $E = E_{F} - 0.90\,eV$ in a $k_x$-$k_y$ plane centered at the $\Gamma$ point~\cite{singhPRB2017_Rashba}. Color depicts the spin projection. 2D spin texture plots reveal the presence of a Rashba type spin-splitting (linear in $k$) of electronic bands near the $\Gamma$ point in BiSb monolayer. Warping effects arising due to the $k^3$ terms in the lower energy valence bands can also be noticed in Fig.~\ref{fig:2DST1}(b). }  
\label{fig:2DST1}
\end{figure}

For example, in Fig.~\ref{fig:2DST1} we plot $S_x$, $S_y$, and $S_z$ spin projections on the 2D spin texture of BiSb monolayer, which is a Rashba semiconductor (Rashba spin-splitting takes place the $\Gamma$ point of BZ), computed in a $k_{x}-k_{y}$ mesh centered at the $\Gamma$ point. Fig.~\ref{fig:2DST1}(a) shows spin-texture calculated above Fermi-energy ($E_{F}$) at constant energy value $E = E_{F} + 0.60\,eV$ (conduction bands), and Fig.~\ref{fig:2DST1}(a) shows the spin-texture calculated below Fermi surface at constant energy value $E = E_{F} - 0.90\,eV$ (valence bands). One can notice linear in $k$ Rashba spin-splitting effects in Fig.~\ref{fig:2DST1}(a), and additional warping effects in Fig.~\ref{fig:2DST1}(b) appearing due to the higher order $k^3$ terms in lower energy valence bands.  

\begin{figure}[htpb!]
  \includegraphics[width=14cm, keepaspectratio=true]{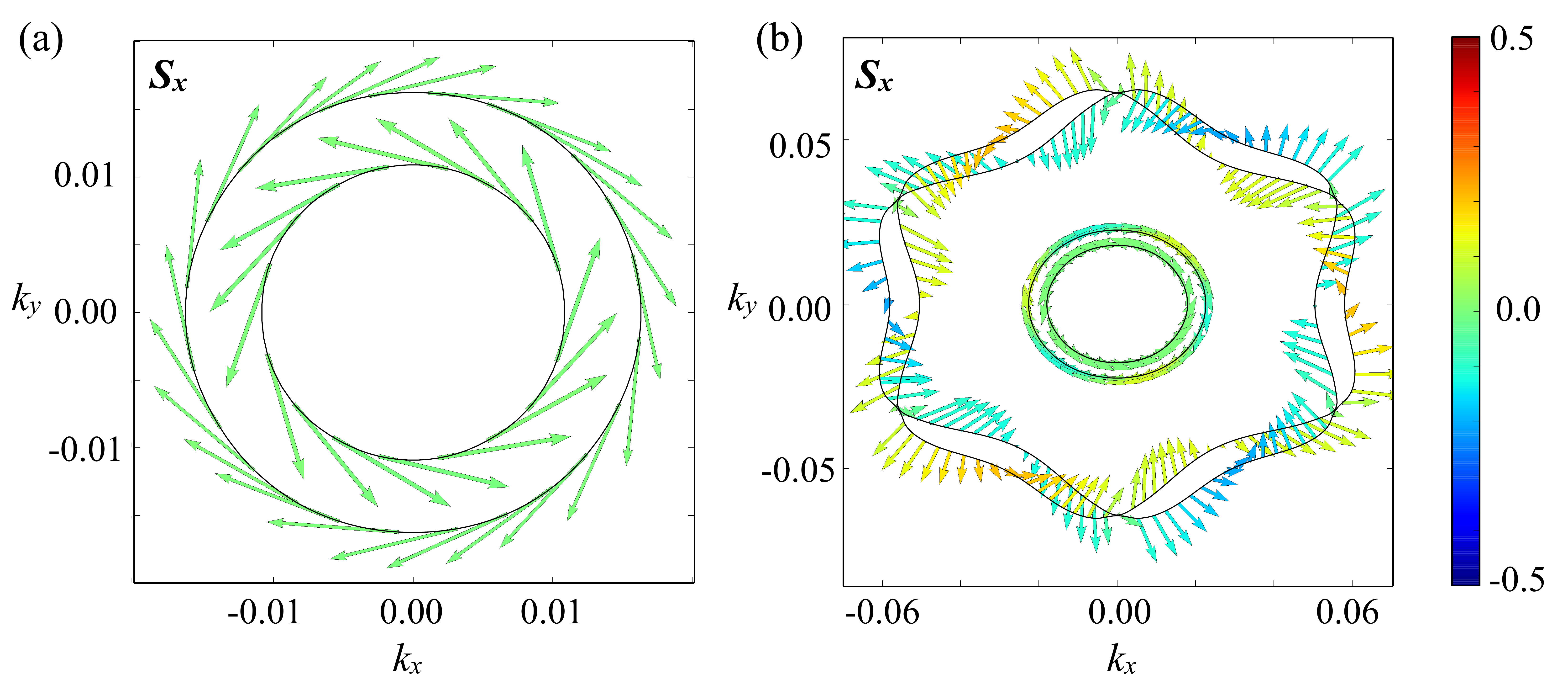}
  \caption{ Projection of $S_x$ spin component shown using arrows instead of heat map as in Fig.~\ref{fig:2DST1}. Spin texture computed at a constant energy surface (a) $E = E_{F} + 0.60\,eV$, and (b) $E = E_{F} - 0.90\,eV$. All other details are same as in Fig.~~\ref{fig:2DST1}. }
    \label{fig:2DST2}
\end{figure}

One could also plot spin texture using arrows instead of a heat map, as shown in Fig.~\ref{fig:2DST2}. This can be done by setting tag: \textbf{noarrow}=False. 

To set maximum and minimum energy values for color map, one could use \texttt{vmin} and \texttt{vmax} tags. For example, \texttt{vmin=-0.5}, \texttt{vmax=0.5} in Fig.~\ref{fig:2DST1} and Fig.~\ref{fig:2DST2}.

\subsection{Identification of Weyl points}
Weyl points appear in the momentum space when two spin non-degenerate electronic bands (one valence and one conduction) linearly cross or touch each other near the Fermi-level forming a topologically non-trivial gapless point in the energy-momentum space. Depending upon the dispersion of valence and conduction bands, Weyl points can be categorized into two types: (i) Type-I, when Weyl cone is not titled and the crossing bands have opposite Fermi-velocity~\cite{SoluyanovNature2015, BinghaiReview2017, winkler2019topology}, and (ii) type-II, when Weyl cone is titled such that Weyl points appear at the touching point of an electron and a hole pocket. In the later case, the crossing bands have unidirectional Fermi-velocity with different magnitudes. When such band crossing occurs between spin degenerate bands, thus formed gapless points are referred as Dirac points. Analogous to Weyl points, Dirac points can also be classified into two type-I and type-II categories. 

\begin{figure}[htpb!]
  \includegraphics[width=14cm, keepaspectratio=true]{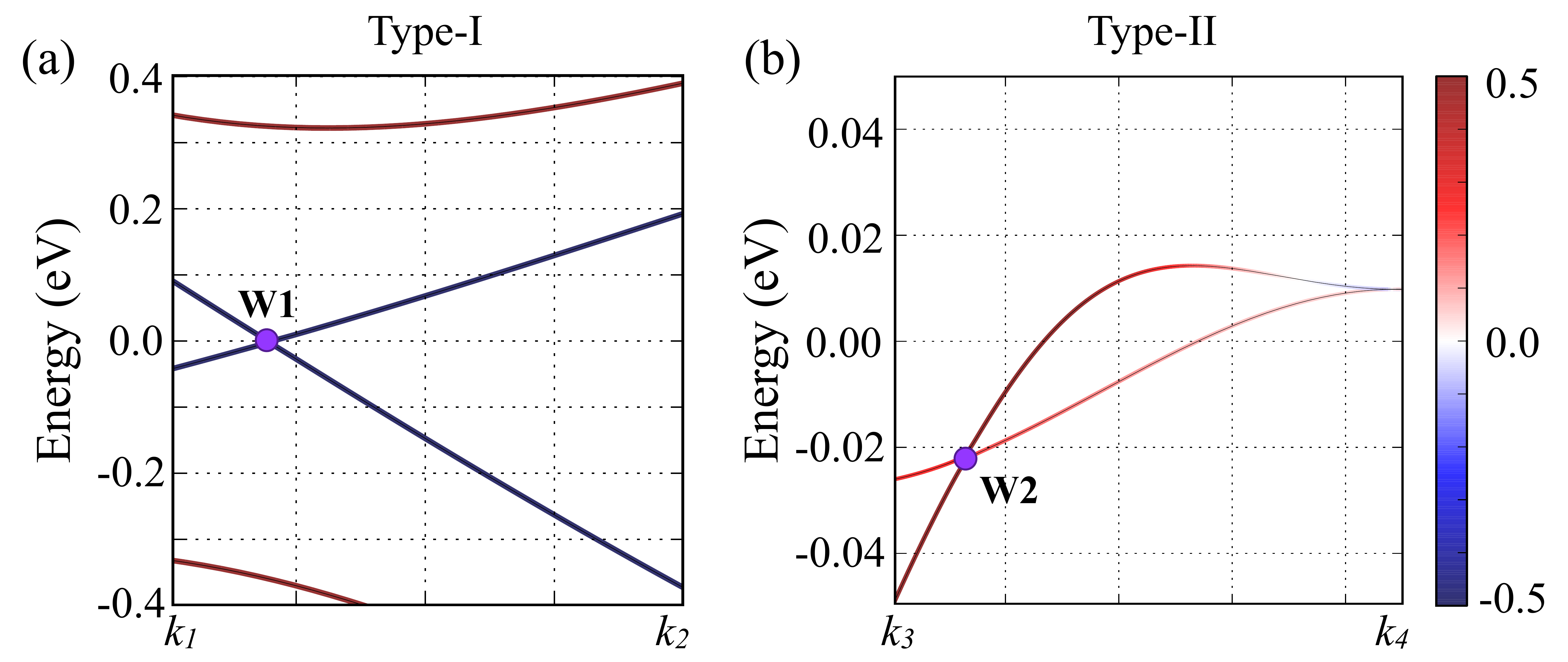}
  \caption{Projection of one of the spin components ($S_x$) on the bandstructure of TaSb near the type-I (a) and type-II (b) Weyl points~\cite{singhPRM2018}. Spin non-degenerate bands with linear dispersion forming a gapless point near the Fermi-level can be observed. Color depicts the spin projection on electronic bands. Direct coordinates of $k_1$, $k_2$, $k_3$, and $k_4$ are (0.5, -0.25, 0.0421), (0.430, -0.215, 0.0421), (-0.040, -0.040, 0.357), and (0.000, 0.000, 0.357), respectively. }  
\label{fig:WPs}
\end{figure}

We can use the PyProcar code to check the spin degeneracy of electronic bands near the crossing points and determine the type of gapless point by analyzing the dispersion of bands in the vicinity of the  crossing point. Fig.~\ref{fig:WPs} shows the spin projected bandstructure of TaSb, which is a topological metal hosting both type-I and type-II Weyl points. For more details regarding the topological properties of TaSb and other similar topological metals, we refer the reader to Refs.~\cite{singhPRM2018, winkler2019topology}.

\subsection{Compare Bands} \label{sec:comparebands}
\begin{figure}[h!]
  \includegraphics[scale=0.6]{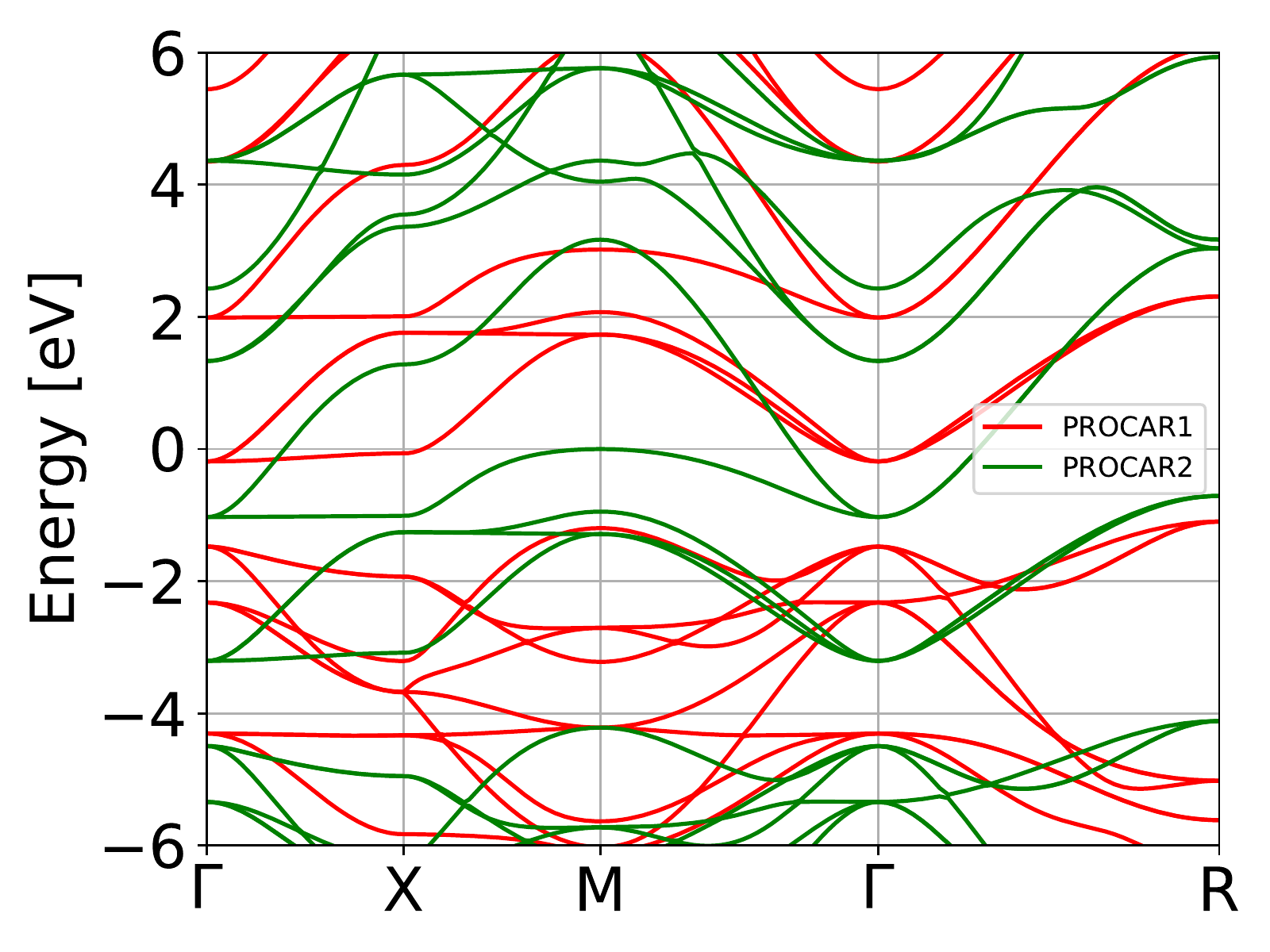}
  \caption{The comparison of energy bands for two systems of SrVO$_3$ with a shift in their Fermi energy}  
  \label{fig:bandscompare}
\end{figure}

This module is useful to compare different bands from different materials on the same band plot (Fig.~\ref{fig:bandscompare}). The bands are plotted for the same $k$-path in order to have a meaningful comparison but they do not need to have the same number of $k$-points in each interval. The \texttt{bandscompare()} function contains all the parameters that are used in the \texttt{bandsplot()} along with an added feature of displaying a \textbf{legend} to help differentiate between the two different band structures. Different \textbf{marker} styles can be used as well. 

\bigskip
\noindent 
Usage:
\begin{lstlisting}
pyprocar.bandscompare(`PROCAR1',`PROCAR2',outcar=`OUTCAR1',outcar2=`OUTCAR2',cmap=`seismic',mode=`parametric',marker=`*',marker2=`-.',elimit=[-5,5],kpointsfile=`KPOINTS',legend=`PRO1',legend2=`PRO2')
\end{lstlisting}

\subsection{Concatenating multiple calculations}
Multiple PROCAR files from multiple DFT calculations can be combined with this utility. For instance, performing DFT calculations for MgB$_2$ supercells are very computationally expensive to do in a single run. This utility is particularly useful in such cases of large systems, where one can split the bandstructure calculations along different high-symmetry directions in BZ, and then concatenate the PROCAR files for each separate $k$-paths, and finally plot the full bandstructure in a single plot. The following command concatenates the PROCAR files obtained from three separate bandstructure calculations done along $\Gamma$-K, K-M, and M-$\Gamma$ $k$-path in hexagonal Brillouin zone. 

\bigskip
\noindent 
Usage:
\begin{lstlisting}
pyprocar.cat([`PROCAR_G-K',`PROCAR_K-M',`PROCAR_M-G'],`PROCAR_merged')
\end{lstlisting}	

\noindent To concatenate PROCAR's generated from Abinit assuming the files are all in the same directory, use the following command: 

\bigskip
\noindent 
Usage:
\begin{lstlisting}
pyprocar.mergeabinit(`PROCAR_merged')
\end{lstlisting}

\subsection{Band Unfolding}
Often times, we need to perform DFT calculations for a supercell geometry rather than the primitive cell. In such cases the band structure becomes quite sophisticated due to the folding of the BZ, and it is difficult to compare the band structure of supercell with that of the primitive cell. The purpose of the band unfolding scheme \cite{Boykin2005,Ku2010,Popescu2012,Allen2013} is to represent the bands within the primitive cell BZ. By calculating the unfolding weight function~\cite{Allen2013} and plotting the fat bands with the line width proportional to the weight, the unfolded bands can be highlighted. Here we use a $2\times2 \times 2$ MgB$_2$ supercell as an example to show the unfolding of band structure. In the bulk structure, where the primitive cell translation symmetry is preserved, the unfolded band is exactly the same as calculated from primitive cell structure. In a structure with a B atom replaced by Al, the translation symmetry can be seen as approximated, and we can still get an unfolded band, with some band smearing out. By comparing these two, we can clearly see the shifting and smearing of the band. The ASE library \cite{Hjorth_Larsen_2017} was employed to perform atomic and cell manipulations required for the unfolding. 

\begin{figure}[htbp]
  (a)\includegraphics[width=0.45\textwidth]{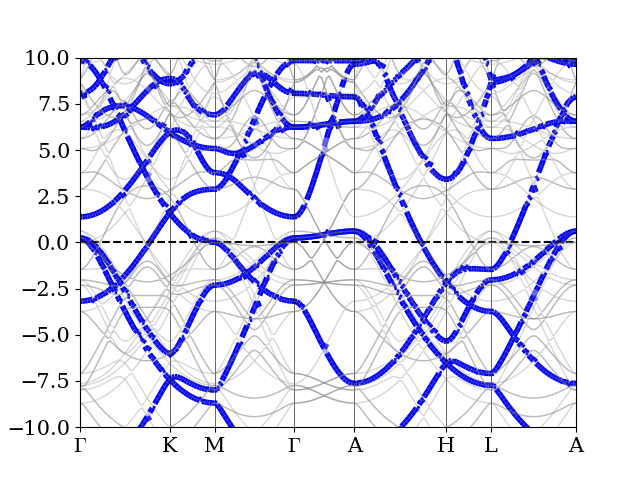}
  (b)\includegraphics[width=0.45\textwidth]{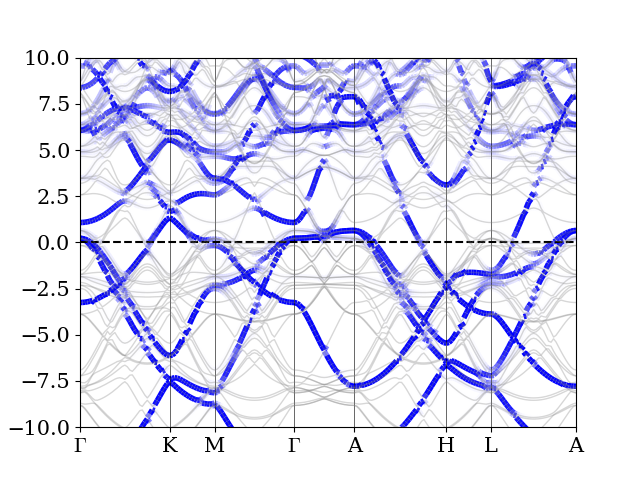}
  \caption{Band structure unfolded into primitive cell BZ of $2 \times 2 \times 2$ supercell (a) MgB$_2$ bulk structure, and (b) MgB$_2$ with one B atom replaced by Al in a $2 \times 2 \times 2$ supercell. The gray lines show the original bands. The width of the blue line denotes the weight of the unfolding. } \end{figure}

\noindent Usage:
First, calculate the band structure in the primitive cell BZ. The PROCAR should be produced with the phase factor included, by setting LORBIT=12 in VASP. Then the unfold module can be used to plot the unfolded band.

\begin{lstlisting}[]
import numpy as np
pyprocar.unfold(
        fname=`PROCAR',
        poscar=`POSCAR',
        outcar=`OUTCAR',
        supercell_matrix=np.diag([2, 2, 2]),
        efermi=None,
        shift_efermi=True,
        elimit=(-5, 15),
        kticks=[0, 36, 54, 86, 110, 147, 165, 199],
        knames=[`$\Gamma$', `K', `M', `$\Gamma$', `A', `H', `L', `A'],
        print_kpts=False,
        show_band=True,
        savefig=`unfolded_band.png')
\end{lstlisting}

\section{Conclusion}
PyProcar is a user friendly, open source Python library that can be easily used for a variety of DFT pre- and post-processing calculations. We have demonstrated its capability through providing examples for each functionality through a set of four materials with unique characteristics. PyProcar's main specialty lies in its ability to project spin, orbitals and atoms in band structures and 2D and 3D Fermi and constant energy surfaces without involving lengthy, complex syntax. The code is freely available to download at \href{https://github.com/romerogroup/pyprocar}{https://github.com/romerogroup/pyprocar}, and via \texttt{pip}. An easy to follow user manual is available at \href{https://romerogroup.github.io/pyprocar/}{https://romerogroup.github.io/pyprocar/}. The PROCAR format is easy to implement in any DFT code, rendering PyProcar accessible across a wide range of DFT codes. We hope this tool will be useful to computational materials scientists in exploring state-of-the-art novel material which would in turn immensely impact the materials community. 
\bigskip

\noindent \emph{Acknowledgements:}
This work used the XSEDE which is supported by National Science Foundation grant number ACI-1053575. XH and EB acknowledge the ARC project AIMED and the F.R.S-FNRS PDR project MaRePeThe (GA 19528980). The authors also acknowledge the support from the Texas Advances Computer Center (with the Stampede2 and Bridges supercomputers),  the PRACE project TheDeNoMo and on the CECI facilities funded by F.R.S-FNRS (Grant No. 2.5020.1) and Tier-1 supercomputer of the F\'ed\'eration Wallonie-Bruxelles funded by the Walloon Region (Grant No. 1117545). This work was supported by the DMREF-NSF 1434897, NSF OAC-1740111 and DOE DE-SC0016176 projects. We acknowledge the West Virginia University supercomputing clusters; Spruce Knob and Thorny Flat which were used for the development of the library. FM acknowledges the support from Fondecyt grants \#1150806, \#1191353, the
Center for the Development of Nanoscience and Nanotechnology CEDENNA FB-0807 and the supercomputing infrastructure of the NLHPC (ECM-02). A special thanks goes to Dr. Guillermo Avendaño Franco for his invaluable support.

\section*{References}
\bibliographystyle{elsarticle-num}
\bibliography{References}
\end{document}